\documentclass[journal, a4paper, twoside]{IEEEtran}
\usepackage{cite}
\usepackage{amsmath,amssymb,amsfonts}
\usepackage{algorithmic}
\usepackage{graphicx}
\usepackage{textcomp}

\begin{document}
\title{Private Coded Computation for Machine Learning}
\author{Minchul Kim, Heecheol Yang, and Jungwoo Lee \\
Department of Electrical and Computer Engineering, Seoul National University, Seoul, Korea \\
E-mail: kmc1222@cml.snu.ac.kr, \{hee2070, junglee\}@snu.ac.kr \vspace{-3mm}
}

\maketitle

\begin{abstract}

In a distributed computing system for the master-worker framework, an erasure code can mitigate the effects of slow workers, also called stragglers.
The distributed computing system combined with coding is referred to as \textit{coded computation}.
We introduce a variation of coded computation that protects the master's privacy from the workers, which is referred to as \textit{private coded computation}.
In private coded computation, the master needs to compute a function  of its own dataset and one of the datasets in a library exclusively shared by the external workers.
After the master recovers the result of the desired function through coded computation, the workers should not know which dataset in the library was desired by the master, which implies that the master's privacy is protected. 
We propose a private coded computation scheme for matrix multiplication, namely  \textit{private polynomial codes}, based on \textit{polynomial codes} for conventional coded computation.
As special cases of private polynomial codes, we propose \textit{private one-shot polynomial codes} and \textit{private asynchronous polynomial codes}.
Whereas the private one-shot polynomial code achieves a lower communication load from the master to each worker, the private asynchronous polynomial code achieves faster computation than private one-shot polynomial codes.
In terms of computation time and communication load, we compare private one-shot polynomial codes and private asynchronous polynomial codes with a conventional robust private information retrieval scheme which can be directly applied to coded computation.
\end{abstract}


\maketitle

\section{Introduction}
\label{intro}
In a distributed computing system where a master partitions a massive computation into smaller sub-computations and distributes these sub-computations to several workers in order to reduce the runtime to complete the whole computation, some slow workers can be bottleneck of the process. 
These slow workers are called \textit{stragglers} and mitigating the effect of these stragglers is one of the major issues in distributed computing. 
Recently, a coding technique was introduced for straggler mitigation \cite{KLee}. 
In \cite{KLee}, for a matrix-vector multiplication, the matrix is $(n,k)$-MDS coded and distributed to $n$ workers so that each encoded matrix is assigned to one worker. 
Each worker multiplies the coded submatrix by a vector and returns the multiplication to the master. 
After $k$ out of $n$ workers return their multiplications, the master can decode the whole computation. 
Since the computation of the slowest $n-k$ workers is ignored, at most $n-k$ stragglers can be mitigated. 
This kind of approach to distributed computing is referred to as \textit{coded computation}. 
Several follow-up studies of coded computation were proposed \cite{followup1}-\cite{followup5}. 

In \cite{PC}, \textit{polynomial codes} were proposed for matrix multiplication, where the minimum number of workers for the master to decode the whole computation does not depend on the number of workers. 
In \cite{Strag1}, a coded computation scheme that sub-blocks a MDS code into small blocks was proposed. 
In this scheme, several small blocks are assigned to each worker. Each worker processes its assigned blocks sequentially, block-by-block, and transmits the partial per-block results to the master. 
Since the size of each block is small enough for the stragglers to compute, all of the workers contribute to the computation, thus reducing the runtime. 
However, after processing their assigned blocks, faster workers stop working and must wait for slower workers to process their blocks. 
If these faster workers could continue working throughout the coded computation process, the runtime would be further reduced. 
Let us call a coded computation scheme where all of the workers continue working until the completion of the coded computation process by \textit{asynchronous coded computation}.

In this paper, we introduce a variation of coded computation that protects the master's privacy from the workers, which is referred to as \textit{private coded computation}.
In the private coded computation, the master requires distributed computing on a function $f$ of its own data $\mathbf{A}$ and specific data $\mathbf{B}_D$ included in a library $\mathbf{B}$, which is exclusively shared by external workers.
For each worker, the master encodes $\mathbf{A}$ with an encoding function $g^D_{\mathbf{A}}$, sends encoded data to the worker, and requests the worker to encode $\mathbf{B}$ with an encoding function $g^D_{\mathbf{B}}$ and compute a function $f_{\text{W}}(g^D_{\mathbf{A}}(\mathbf{A}),g^D_{\mathbf{B}}(\mathbf{B}))$. 
After the master recovers the result of desired function $f(\mathbf{A},\mathbf{B}_D)$ from the computation results of $f_{\text{W}}$ returned by the workers, the workers should not be able to identify that $\mathbf{B}_D$ is desired by the master, which would imply that the master's privacy is protected.
Private coded computation will be explained in further detail in Section \ref{system}.

As a motivating example of the private coded computation, we may consider a user who employs an artificial intelligence (AI) assistant, e.g. Google Assistant or Siri, with its mobile.
We assume that the user can request a recommendation from an AI assistant of an item which is included in one of $M$ categories that the AI assistant can recommend, e.g. movies, games, restaurants, and so on.
We refer to the $M$ categories as a library $\mathbf{B}$ and denote them by $\{\mathbf{B}_k\}_{k=1}^M$ such that $\mathbf{B}=\{\mathbf{B}_k\}_{k=1}^M$.  
We also assume that the user stores its preference parameter $\mathbf{A}$.
When the user requests a recommendation from the AI assistant of an item in a category $\mathbf{B}_D$, the assistant encodes $\mathbf{A}$ and sends encoded data to several distributed workers, e.g. data centers, for recovering $f(\mathbf{A},\mathbf{B}_D)$ in a distributed way.
After recovery, the AI assistant can decide the recommended item based on $f(\mathbf{A},\mathbf{B}_D)$.
We assume that the user can delete the recommendation service usage record right after the item is recommended so that the AI assistant does not identify the user's recommendation service usage pattern.

Generally, the user uses this recommendation service according to its life cycle.
That is, if the workers track the recommendation service usage records, the user's life cycle is revealed to them, which implies that the user's privacy has been invaded.
We remark that this privacy invasion on the user's life cycle is related to $\mathbf{B}$, not $\mathbf{A}$.
That is, encrypting the user's preference parameter $\mathbf{A}$ cannot protect the user's privacy on the life cycle.
Therefore, in order to protect the user's privacy, the workers should not know that a particular $\mathbf{B}_D$ is desired by a user, which motivates the private coded computation.

\begin{figure*}
    \centerline{\includegraphics[width=16cm]{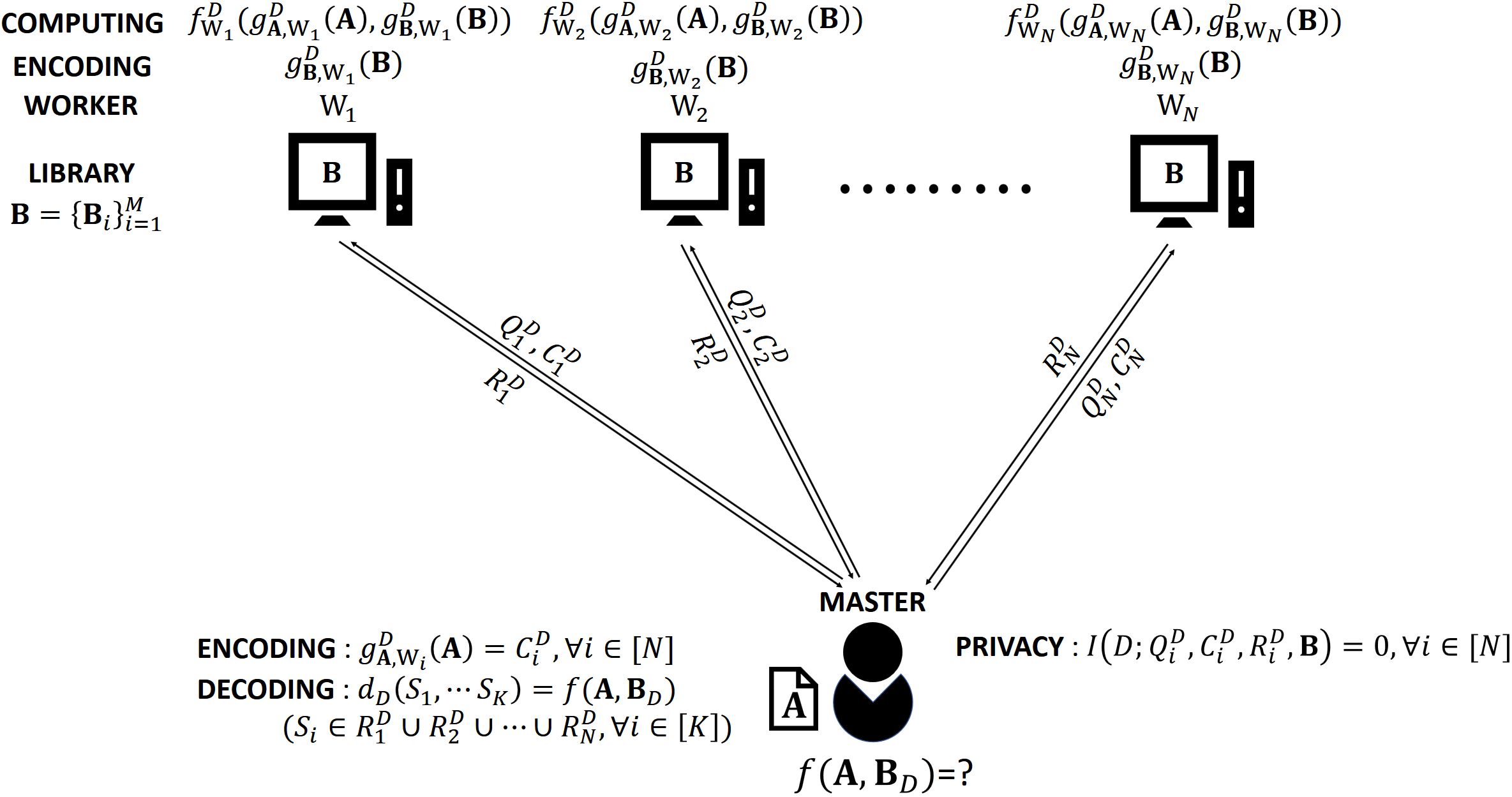}}
    \caption{
    The overall process of private coded computation. 
    Using $N$ workers $\{\text{W}_i\}_{i=1}^N$, a master wants to compute a function $f$ of its own dataset $\mathbf{A}$ and a desired dataset $\mathbf{B}_D$ in a library $\mathbf{B}$ of $M$ datasets $\{\mathbf{B}_k\}_{k=1}^M$, which is shared by the workers, while the workers do not identify the desired index $D$. 
    The master encodes $\mathbf{A}$ for each worker $\text{W}_i$ with an encoding function $g^D_{\mathbf{A},\text{W}_i}$ and transmits encoded data $C_i^D$ to $\text{W}_i$. 
    The master also sends the queries $Q_i^D$ to $\text{W}_i$ for requesting $\text{W}_i$ to encode $\mathbf{B}$ with its encoding function $g^D_{\mathbf{B},\text{W}_i}$ and to compute a function $f^D_{\text{W}_i}(g^D_{\mathbf{A},\text{W}_i}(\mathbf{A}), g^D_{\mathbf{B},\text{W}_i}(\mathbf{B}))$. 
    Each worker $\text{W}_i$ sequentially computes the function $f^D_{\text{W}_i}(g^D_{\mathbf{A},\text{W}_i}(\mathbf{A}), g^D_{\mathbf{B},\text{W}_i}(\mathbf{B}))$ and returns sub-computation results of $f^D_{\text{W}_i}$, $R_i^D$, to the master. 
    After receiving $K$ sub-computation results $\{S_i\}_{i=1}^K$ from $N$ workers, the master can decode the sub-computation results with decoding function $d_D$ to recover $f(\mathbf{A},\mathbf{B}_D)$. 
    For each $\text{W}_i$, the constraint for the master's privacy is given by $I(D;Q_i^D,C_i^D,R_i^D,\mathbf{B})=0$.  
    }
    \label{overall}
    \vspace{0mm}
\end{figure*}

In this paper, we propose a private coded computation scheme for matrix multiplication, based on \textit{polynomial codes} in \cite{PC}. 
We refer to this scheme as \textit{private polynomial codes}.
In order to protect the master's privacy, the workers are grouped according to the assigned work in the private polynomial codes,  which is different from the conventional polynomial codes. 

As special cases of private polynomial codes, we propose \textit{private one-shot polynomial codes} and \textit{private asynchronous polynomial codes}.
In the private one-shot polynomial codes, the whole computation is converted into several sub-computations, and only one  sub-computation is assigned to each worker. 
Since each worker only computes the sub-computation once, we call this scheme private one-shot polynomial codes. 
The computation of stragglers is ignored in private one-shot polynomial codes. 
By contrast, in private asynchronous polynomial codes, the whole computation is converted into much smaller sub-computations so that even the stragglers, with their low computational capability, can finish some sub-computations. 
In private asynchronous polynomial codes, several sub-computations are to be computed by each worker. 
If assigned enough sub-computations, every worker within each group can continue working throughout the coded computation process,  which reduces the computation time for coded computing. 

Whereas the private asynchronous polynomial codes achieve faster computation time, the private one-shot polynomial codes achieve relatively lower communication load from the master to the workers.
In terms of computation time and communication load, we compare private one-shot polynomial codes and private asynchronous polynomial codes with the conventional optimal robust private information retrieval (RPIR) scheme in \cite{RPIR_Sun}. 
Note that the RPIR scheme in \cite{RPIR_Sun} is directly applicable for coded computation.

\textit{Notation }: We use $[N]$ to denote a set comprised of $N$ elements, 1 to $N$. A set comprised of $M$ elements, $N+1$ to $N+M$ is denoted by $[N+1:N+M]$.

\section{System model}
\label{system}
In this section, we describe a system model of private coded computation.
There is a master who has its own dataset $\mathbf{A}$, where $\mathbf{A}$ is an element (matrix) in a vector space $\mathbb{V}_1$ over a field $\mathbb{F}$.
There are also $N$ external workers $\{\text{W}_i\}_{i=1}^N$, and these workers share a library $\mathbf{B}$ which  consists of $M$ different datasets $\{\mathbf{B}_k\}_{k=1}^M$.
Each dataset $\mathbf{B}_k$ is an element (matrix) in a vector space $\mathbb{V}_2$ over the same field $\mathbb{F}$.
The master needs distributed computing on a function $f$ of $\mathbf{A}$ and one of $M$ datasets $\{\mathbf{B}_k\}_{k=1}^M$ in library $\mathbf{B}$, where $f:(\mathbb{V}_1,\mathbb{V}_2) \rightarrow \mathbb{V}_3$ for a vector space $\mathbb{V}_3$ over the same field $\mathbb{F}$.
We denote the desired dataset by $\mathbf{B}_D$.
Therefore, the whole computation desired by the master is denoted by $f(\mathbf{A},\mathbf{B}_D)$. 
Since we consider private coded computation for matrix multiplication, $f(\mathbf{A},\mathbf{B}_D)=\mathbf{AB}_D$ in this paper.

The whole computation is converted into several sub-computations and the master assigns one or more sub-computations to each worker.
We assume that the same number of sub-computations are assigned to each worker and that each worker returns each sub-computation result to the master whenever the worker finishes a sub-computation.
When sufficient number of sub-computation results are returned to the master, the master can recover the whole computation $f(\mathbf{A},\mathbf{B}_D)$ based on the received sub-computation results. 
We denote the minimum number of sub-computation results to recover $f(\mathbf{A},\mathbf{B}_D)$ by $K$.
If only one sub-computation is assigned to each worker, the sub-computation results returned by the fastest $K$ workers are enough to recover the whole computation.
That is, the minimum number of sub-computation results to recover $f(\mathbf{A},\mathbf{B}_D)$ equals the minimum number of workers to recover $f(\mathbf{A},\mathbf{B}_D)$, which was defined as \textit{recovery threshold} in \cite{PC}. 
In this case, the slowest $N-K$ workers become stragglers, since they do not finish their sub-computations and thus do not contribute to the whole computation.
If several sub-computations are assigned to each worker, the master can recover $f(\mathbf{A},\mathbf{B}_D)$ when $K$ sub-computation results are returned across $N$ workers, including the computation of slow workers.
After the master recovers the whole computation, each worker should not be able to identify that $\mathbf{B}_D$ is desired by the master, thus protecting the master's privacy.
In this paper, we assume that the workers do not collude with each other in that each worker does not know which sub-computations are assigned to, computed by, and returned by the other workers.

The master's own dataset $\mathbf{A}$ and the library $\mathbf{B}$ are encoded for the private coded computation.
Note that the master's own dataset $\mathbf{A}$ is encoded by the master whereas the library $\mathbf{B}$ is encoded by each worker.
The master encodes $\mathbf{A}$ for each worker $\text{W}_i$.
We denote the encoding function of $\mathbf{A}$ for the worker $\text{W}_i$ and desired matrix $\mathbf{B}_D$ by $g^D_{\mathbf{A},\text{W}_i}$, where $g^D_{\mathbf{A},\text{W}_i}:\mathbb{V}_1 \rightarrow \mathbb{U}_1^{\delta_1}$ for a vector space $\mathbb{U}_1$ over the same field $\mathbb{F}$.
Note that the parameter $\delta_1$ is related to the number of sub-computations assigned to each worker.
The master sends the encoded data $g^D_{\mathbf{A},\text{W}_i}(\mathbf{A})$ to the worker $\text{W}_i$ and also sends the queries for requesting $\text{W}_i$ to encode the library $\mathbf{B}$.
We denote the encoding function of the worker $\text{W}_i$ for the library $\mathbf{B}=\{\mathbf{B}_k\}_{k=1}^M$ and the desired dataset $\mathbf{B}_D$ by $g^D_{\mathbf{B},\text{W}_i}$, where $g^D_{\mathbf{B},\text{W}_i}:\mathbb{V}_2^M \rightarrow \mathbb{U}_2^{\delta_2}$ for a vector space $\mathbb{U}_2$ over the same field $\mathbb{F}$.
The parameters $\delta_1$ and $\delta_2$ are jointly related to the number of sub-computations assigned to each worker.
The master also sends the queries to the worker $\text{W}_i$ to compute a function of $g^D_{\mathbf{A},\text{W}_i}(\mathbf{A})$ and $g^D_{\mathbf{B},\text{W}_i}(\mathbf{B})$ and return the computation result of the function to the master.
That is, the worker $\text{W}_i$ computes $f^D_{\text{W}_i}(g^D_{\mathbf{A},\text{W}_i}(\mathbf{A}), g^D_{\mathbf{B},\text{W}_i}(\mathbf{B}))$.
We denote the function of $\text{W}_i$ by $f^D_{\text{W}_i}:(\mathbb{U}_1^{\delta_1},\mathbb{U}_2^{\delta_2})\rightarrow \mathbb{U}_3^{\delta_3}$ for a vector space $\mathbb{U}_3$ over the same field $\mathbb{F}$.
Note that the parameter $\delta_3$ denotes the number of sub-computations to be returned by the worker $\text{W}_i$ and $\delta_3$ is the same across workers.
As was previously assumed, if $\delta_3$ is larger than 1, the worker $\text{W}_i$ sequentially computes the function $f^D_{\text{W}_i}$ one by one, and returns each sub-computation result to the master upon finishing each sub-computation.
Without considering where the sub-computation result comes from, we denote the $i$th sub-computation result returned to the master by $S_i$, where $S_i$ is an element in the vector space $\mathbb{U}_3$. 
After $K$ sub-computation results $\{S_i\}_{i=1}^K$ across the $N$ workers are returned to the master, the master can recover the whole computation $f(\mathbf{A},\mathbf{B}_D)$ by decoding $\{S_i\}_{i=1}^K$.
If we denote the decoding function at the master by $d_D:\mathbb{U}_3^K\rightarrow \mathbb{V}_3$, the decoding function $d_D$ should satisfy the constraint given by $d_D(S_1, S_2, \cdots , S_K)=f(\mathbf{A},\mathbf{B}_D)$. 

The master's privacy is protected when none of the workers can identify index $D$ of the desired dataset $\mathbf{B}_D$ after the master recovers the whole computation. Since the privacy we consider is information-theoretic privacy, the privacy constraint for each worker $\text{W}_i$ can be expressed as 
\begin{align}
 I(D;Q_i^D,g^D_{\mathbf{A},\text{W}_i}(\mathbf{A}),f^D_{\text{W}_i}(g^D_{\mathbf{A},\text{W}_i}(\mathbf{A}), g^D_{\mathbf{B},\text{W}_i}(\mathbf{B})),\mathbf{B})=0, \nonumber
\end{align}
where $Q_i^D$ denotes the queries that the master sends to the worker $\text{W}_i$ for encoding $g^D_{\mathbf{B},\text{W}_i}(\mathbf{B})$ and computing $f^D_{\text{W}_i}(g^D_{\mathbf{A},\text{W}_i}(\mathbf{A}), g^D_{\mathbf{B},\text{W}_i}(\mathbf{B}))$.

For a simpler expression, we denote $g^D_{\mathbf{A},\text{W}_i}(\mathbf{A})$ and $f^D_{\text{W}_i}(g^D_{\mathbf{A},\text{W}_i}(\mathbf{A}), g^D_{\mathbf{B},\text{W}_i}(\mathbf{B}))$ by $C_i^D$ and $R_i^D$, respectively, so that the privacy constraint becomes
\begin{align}
 I(D;Q_i^D,C_i^D,R_i^D,\mathbf{B})=0. 
\label{privacyconstraint}
\end{align}
The overall process of the private coded computation is depicted in Fig.1.

\section{Private polynomial codes}
In this section, we propose private polynomial codes for matrix multiplication.
We describe the scheme with two illustrative examples and generally describe the private polynomial codes. 
We also show that the master's privacy is protected.
Finally, we explain certain special cases of private polynomial codes, which are referred to as private one-shot polynomial codes and private asynchronous polynomial codes.

\subsection{Illustrative examples}
\label{example}
As the first example, we assume that the master has a matrix 
$\mathbf{A}\in \mathbb{F}_q^{r\times s}$ 
for sufficiently large finite field 
$\mathbb{F}_q$ 
and that there are  12 non-colluding workers 
$\{\text{W}_{n}\}_{n=1}^{12}$ 
where each worker has a library of two matrices 
$\mathbf{B}_{1},\mathbf{B}_{2}\in \mathbb{F}_q^{s\times t}$.
As in Section \ref{system}, we denote the library by $\mathbf{B}$.
Let us assume that the master wants to compute 
$\mathbf{AB}_1$ 
using 
$\{\text{W}_{n}\}_{n=1}^{12}$
while hiding that the master desires 
$\mathbf{B}_{1}$ 
from the workers. 
The matrix $\mathbf{A}$ can be partitioned into two submatrices 
$\mathbf{A}_{0}, \mathbf{A}_{1}\in \mathbb{F}_q^{r/2\times s}$ 
so that 
$\mathbf{A}$=$\begin{bmatrix}
{\mathbf{A}_0} \\ {\mathbf{A}_1}
\end{bmatrix}$ and each of $\mathbf{B}_1,\mathbf{B}_2$ 
are partitioned into two submatrices 
$\mathbf{B}_{k,1}, \mathbf{B}_{k,2}\in \mathbb{F}_q^{s\times t/2}, k\in[2]$
, so that 
$\mathbf{B}_k$=$\begin{bmatrix}
{\mathbf{B}_{k,1}} & {\mathbf{B}_{k,2}}
\end{bmatrix}$. 
Therefore, 
$\mathbf{AB}_1=\begin{bmatrix}
\mathbf{A}_0\mathbf{B}_{1,1} & \mathbf{A}_0\mathbf{B}_{1,2} \\ \mathbf{A}_1\mathbf{B}_{1,1} & \mathbf{A}_1\mathbf{B}_{1,2}
\end{bmatrix}$. 
The polynomial codes for $\mathbf{A}$,  
$\mathbf{B}_{1}$ and $\mathbf{B}_{2}$
are as follows.
\begin{gather}
\mathbf{\tilde A}(x)=\mathbf{A}_0+\mathbf{A}_1 x, \:\:\:\: \mathbf{\tilde B}_k(y)=\mathbf{B}_{k,1} y+\mathbf{B}_{k,2} y^{2},
\end{gather}
where $k\in[2]$ and $x,y \in \mathbb{F}_q$ denote the variables of polynomials $\mathbf{\tilde A}$ and $\mathbf{\tilde B}_k$, respectively. 

Note that the master's own matrix $\mathbf{A}$ and the library $\mathbf{B}$ are encoded separately, which differs from the conventional polynomial codes.
We denote the evaluations of 
$\mathbf{\tilde A}$ at $x=x_i$ and $\mathbf{\tilde B}_k$ at $y=y_i$
by 
$\mathbf{\tilde A}(x_i)$ and $\mathbf{\tilde B}_k(y_i)$
, respectively. 
For the desired matrix $\mathbf{B}_1$ and each worker $\text{W}_i$, the master evaluates $\mathbf{\tilde A}$ at a randomly chosen point $x_{\text{W}_i}$ and sends the evaluation $\mathbf{\tilde A}(x_{\text{W}_i})$ to the worker $\text{W}_i$.
That is, $g^1_{\mathbf{A},\text{W}_i}(\mathbf{A})=\mathbf{\tilde A}(x_{\text{W}_i})$. 
We assume that the points $\{x_{\text{W}_i}\}_{i=1}^{12}$ are distinct from each other.
Since the master sends only one evaluation $\mathbf{\tilde A}(x_{\text{W}_i})$ to each worker, $\delta_1=1$ in the encoding function $g^1_{\mathbf{A},\text{W}_i}$.
The master also sends the queries $Q_i^1$ that request $\text{W}_i$ to encode the library $\mathbf{B}$ with and encoding function $g^1_{\mathbf{B},\text{W}_i}$ and compute a function $f_{\text{W}_i}^1(g^1_{\mathbf{A},\text{W}_i}(\mathbf{A}),g^1_{\mathbf{B},\text{W}_i}(\mathbf{B}))$.

The library $\mathbf{B}$ is encoded as follows.
Firstly, for evaluating $\mathbf{\tilde B}_1$, the workers are divided into three equal-sized groups according to the evaluating points.
For simplicity, we assume that the evaluations points for the desired matrix $\mathbf{\tilde B}_1$ are$y=y_1$ for the workers $\{\text{W}_i\}_{i=1}^4$, $y=y_2$ for the workers $\{\text{W}_i\}_{i=5}^{8}$, and $y=y_3$ for the workers $\{\text{W}_i\}_{i=9}^{12}$, respectively.
Note that the points $\{y_i\}_{i=1}^3$ are randomly chosen in $\mathbb{F}_q$ and distinct from each other.
We denote the workers $\{\text{W}_i\}_{i=1}^4$, $\{\text{W}_i\}_{i=5}^8$, and $\{\text{W}_i\}_{i=9}^{12}$ by groups $G_1$, $G_2$ and $G_3$, respectively.
Since the workers do not collude with each other, they do not know that they are grouped according to the evaluating points for $\mathbf{\tilde B}_1$. 
Secondly, for all of the workers, the undesired matrix $\mathbf{\tilde B}_2$ is evaluated at a randomly chosen point $y_{4}$ which is distinct from points $\{y_i\}_{i=1}^3$.
Since the workers do not collude with each other, they cannot notice that  $\mathbf{\tilde B}_2$ is evaluated at an identical point $y_{4}$ across workers.
Finally, for each worker $\text{W}_i$ in each group $G_t$, the encoded library is given by $g^1_{\mathbf{B},\text{W}_i}(\mathbf{B})=\mathbf{\tilde B}_1(y_t)+\mathbf{\tilde B}_2(y_4)$.
Since the desired matrix $\mathbf{B}_1$ is evaluated at only one point for each worker, $\delta_2=1$ in the encoding function $g^1_{\mathbf{B},\text{W}_i}$.

After encoding the library, each worker $\text{W}_i$ in each group $G_t$ computes a function $f_{\text{W}_i}^1(g^1_{\mathbf{A},\text{W}_i}(\mathbf{A}),g^1_{\mathbf{B},\text{W}_i}(\mathbf{B}))=\mathbf{\tilde A}(x_{\text{W}_i})(\mathbf{\tilde B}_1(y_t)+\mathbf{\tilde B}_2(y_4))$. 
Since every worker computes only one sub-computation, $\delta_3=1$ in the function $f_{\text{W}_i}^1$.
We can express the sub-computation $\mathbf{\tilde A}(x_{\text{W}_i})(\mathbf{\tilde B}_1(y_t)+\mathbf{\tilde B}_2(y_4))$ as follows.
\begin{flalign}
&\mathbf{\tilde A}(x_{\text{W}_i})(\mathbf{\tilde B}_1(y_t)+\mathbf{\tilde B}_2(y_4))\nonumber&& \\ &=(\mathbf{A}_0+\mathbf{A}_1x_{\text{W}_i})(\mathbf{B}_{1,1} y_t+\mathbf{B}_{1,2} y_t^{2}+\mathbf{B}_{2,1} y_{4}+\mathbf{B}_{2,2} y_{4}^{2})&& \nonumber \\ \nonumber &=\mathbf{Z}_{t,0}+\mathbf{Z}_{t,1} x_{\text{W}_i},&&  \nonumber 
\end{flalign}
where $\mathbf{Z}_{t,0}$ and $\mathbf{Z}_{t,1}$ are given by
\begin{gather}
\mathbf{Z}_{t,0}=\mathbf{A}_0(\mathbf{B}_{1,1} y_t+\mathbf{B}_{1,2} y_t^{2}+\mathbf{B}_{2,1} y_{4}+\mathbf{B}_{2,2} y_{4}^{2}), \nonumber \\ 
\mathbf{Z}_{t,1}=\mathbf{A}_1(\mathbf{B}_{1,1} y_t+\mathbf{B}_{1,2} y_t^{2}+\mathbf{B}_{2,1} y_{4}+\mathbf{B}_{2,2} y_{4}^{2}). \nonumber  
\end{gather}

Since the evaluating points $y_{t}$ and $y_{4}$ are identical across workers in group $G_t$, we can write the polynomial of $x$ to be decoded by the master as $\mathbf{\tilde A}(x)(\mathbf{\tilde B}_1(y_t)+\mathbf{\tilde B}_2(y_4))=\mathbf{Z}_{t,0}+\mathbf{Z}_{t,1} x$.
Since the degree of polynomial $\mathbf{\tilde A}(x)(\mathbf{\tilde B}_1(y_t)+\mathbf{\tilde B}_2(y_4))$ is 1 and the evaluating points $\{x_{\text{W}_i}\}_{i=1}^{12}$ are distinct from each other, the master can decode the polynomial from the sub-computation results returned by the two fastest workers in $G_t$, by polynomial interpolation with respect to the variable $x$.

We denote the decoding function by $d_D$ and the $i$th sub-computation result returned to the master within group $G_t$ by $S_{t,i}$.
The master can decode the each of the three polynomials $\{\mathbf{\tilde A}(x)(\mathbf{\tilde B}_1(y_t)+\mathbf{\tilde B}_2(y_4))\}_{t=1}^3$ from each set of two sub-computation results $\{S_{t,1},S_{t,2}\}_{t=1}^3$ returned by each of the three groups $\{G_t\}_{t=1}^3$, so that the coefficients $\{\mathbf{Z}_{t,0},\mathbf{Z}_{t,1}\}_{t=1}^{3}$ are obtained.
The coefficients $\{\mathbf{Z}_{t,0}\}_{t=1}^3$ and $\{\mathbf{Z}_{t,1}\}_{t=1}^3$ are evaluations of polynomials $(\mathbf{A}_0\mathbf{B}_{2,1} y_4+\mathbf{A}_0\mathbf{B}_{2,2} y_4^2)+\mathbf{A}_0\mathbf{B}_{1,1} y+\mathbf{A}_0\mathbf{B}_{1,2} y^2$ and $(\mathbf{A}_1\mathbf{B}_{2,1} y_4+\mathbf{A}_1\mathbf{B}_{2,2} y_4^2)+\mathbf{A}_1\mathbf{B}_{1,1} y+\mathbf{A}_1\mathbf{B}_{1,2} y^2$ at $y=\{y_t\}_{t=1}^3$, respectively.
Note that the terms $\mathbf{A}_0\mathbf{B}_{2,1} y_4+\mathbf{A}_0\mathbf{B}_{2,2} y_4^2$ and $\mathbf{A}_1\mathbf{B}_{2,1} y_4+\mathbf{A}_1\mathbf{B}_{2,2} y_4^2$ are constant terms in each polynomial, respectively.
For each $l\in[0,1]$, the master can decode each polynomial $(\mathbf{A}_l\mathbf{B}_{2,1} y_4+\mathbf{A}_l\mathbf{B}_{2,2} y_4^2)+\mathbf{A}_l\mathbf{B}_{1,1} y+\mathbf{A}_l\mathbf{B}_{1,2} y^2$ from $\{\mathbf{Z}_{t,l}\}_{t=1}^3$ with polynomial interpolation with respect to $y$ so that the coefficients $\{\mathbf{A}_l\mathbf{B}_{1,i}\}_{i=1}^{2}$ are obtained.
After decoding all of the polynomials $\{(\mathbf{A}_l\mathbf{B}_{2,1} y_4+\mathbf{A}_l\mathbf{B}_{2,2} y_4^2)+\mathbf{A}_l\mathbf{B}_{1,1} y+\mathbf{A}_l\mathbf{B}_{1,2} y^2\}_{l=0}^1$, the whole computation $\mathbf{AB}_1=\begin{bmatrix}
\mathbf{A}_0\mathbf{B}_{1,1} & \mathbf{A}_0\mathbf{B}_{1,2} \\ \mathbf{A}_1\mathbf{B}_{1,1} & \mathbf{A}_1\mathbf{B}_{1,2}
\end{bmatrix}$ can be recovered from the coefficients $\{\mathbf{A}_l\mathbf{B}_{1,i}\}_{(l,i) = (0,1)}^{(1,2)}$.
Therefore, the minimum number of sub-computation results to recover the whole computation $\mathbf{AB}_1$, which was denoted by $K$ in Section \ref{system}, equals 6 and  $d_D(S_{1,1}, S_{1,2}, S_{2,1}, S_{2,2}, S_{3,1}, S_{3,2})=\mathbf{AB}_1$.
Note that the master can recover the whole computation $\mathbf{AB}_1$ after the slowest group among three groups returns two sub-computation results, thus implying that the computation time of the private polynomial code is dependent on the slowest group.

We now explain the second example.
Compared to the first example, the only difference in the second example is the number of submatrices of $\mathbf{A}$, where $\mathbf{A}$ can be partitioned into 100 submatrices $\{\mathbf{A}_{i}\}_{i=0}^{99} \in \mathbb{F}_q^{r/100 \times s}$ so that
$\mathbf{A}=\begin{bmatrix}
{\mathbf{A}_0}^T & {\mathbf{A}_1}^T & \cdots & {\mathbf{A}_{99}^T}
\end{bmatrix}^T$.  
Note that the size of each submatrix $\mathbf{A}_i$ is 50 times smaller than that of the first example. 
Since $\mathbf{B}_k$=$\begin{bmatrix}
{\mathbf{B}_{k,1}} & {\mathbf{B}_{k,2}}
\end{bmatrix}$, the whole computation $\mathbf{AB}_1$ which is desired by the master is given by 

\begin{align}
\mathbf{AB}_1=\begin{bmatrix}
\mathbf{A}_0\mathbf{B}_{1,1} & \mathbf{A}_0\mathbf{B}_{1,2}   \\ \mathbf{A}_1\mathbf{B}_{1,1} & \mathbf{A}_1\mathbf{B}_{1,2}  \\
\vdots & \vdots \\
\mathbf{A}_{99}\mathbf{B}_{1,1} & \mathbf{A}_{99}\mathbf{B}_{1,2} 
\end{bmatrix}.
\label{wholecomputation}
\end{align}

The polynomial codes for $\mathbf{A}$,  
$\mathbf{B}_{1}$ and $\mathbf{B}_{2}$
are given as follows.
\begin{align}
\mathbf{\tilde A}(x)=\sum_{l=0}^{99}\mathbf{A}_l x^{l}, \:\:\:\: \mathbf{\tilde B}_k(y)=\mathbf{B}_{k,1} y+\mathbf{B}_{k,2} y^{2}, 
\end{align}
where $k\in[2]$ and $x,y \in \mathbb{F}_q$ denote the variables of polynomials $\mathbf{\tilde A}$ and $\mathbf{\tilde B}_k$, respectively.

For the desired matrix $\mathbf{B}_1$ and each worker $\text{W}_i$, the master evaluates $\mathbf{\tilde A}$ at 100 distinct and randomly chosen points $\{x_{\text{W}_i,p}\}_{p=1}^{100}$ and sends the evaluations $\{\mathbf{\tilde A}(x_{\text{W}_i,p})\}_{p=1}^{100}$ to the worker $\text{W}_i$.
That is, $g^1_{\mathbf{A},\text{W}_i}(\mathbf{A})=\{\mathbf{\tilde A}(x_{\text{W}_i,p})\}_{p=1}^{100}$. 
Since the master sends 100 evaluations to each worker, $\delta_1=100$ in the encoding function $g^1_{\mathbf{A},\text{W}_i}$.
The master also sends the queries $Q_i^1$ that request $\text{W}_i$ to encode the library $\mathbf{B}$ with and encoding function $g^1_{\mathbf{B},\text{W}_i}$ and compute a function $f_{\text{W}_i}^1(g^1_{\mathbf{A},\text{W}_i}(\mathbf{A}),g^1_{\mathbf{B},\text{W}_i}(\mathbf{B}))$. 
The library $\mathbf{B}$ is encoded in the same way as in the first example.
That is, all of 12 workers are divided into three equal-sized groups according to the evaluating point for $\mathbf{\tilde B}_1$.

After encoding the library, each worker $\text{W}_i$ in each group $G_t$ computes a function $f_{\text{W}_i}^1(g^1_{\mathbf{A},\text{W}_i}(\mathbf{A}),g^1_{\mathbf{B},\text{W}_i}(\mathbf{B}))=\{\mathbf{\tilde A}(x_{\text{W}_{i},p})(\mathbf{\tilde B}_1(y_t)+\mathbf{\tilde B}_2(y_4))\}_{p=1}^{100}$. 
Since every worker computes 100 sub-computations, $\delta_3=100$ in the function $f_{\text{W}_i}^1$.
Since $\delta_3 >1$, each worker sequentially computes the function $f_{\text{W}_i}^1$ and returns each sub-computation $\mathbf{\tilde A}(x_{\text{W}_{i},p})(\mathbf{\tilde B}_1(y_t)+\mathbf{\tilde B}_2(y_4))$ whenever it finishes, as explained in Section \ref{system}.

We can express each sub-computation $\mathbf{\tilde A}(x_{\text{W}_{i},p})(\mathbf{\tilde B}_1(y_t)+\mathbf{\tilde B}_2(y_4))$ as follows.
\begin{flalign}
&\mathbf{\tilde A}(x_{\text{W}_{i},p})(\mathbf{\tilde B}_1(y_t)+\mathbf{\tilde B}_2(y_4))\nonumber&& \\ &=(\sum_{l=0}^{99}{\mathbf{A}_l x_{\text{W}_i,p}^l})(\mathbf{B}_{1,1} y_t+\mathbf{B}_{1,2} y_t^{2}+\mathbf{B}_{2,1} y_{4}+\mathbf{B}_{2,2} y_{4}^{2})&& \nonumber \\ \nonumber &=\sum_{l=0}^{99}{\mathbf{Z}_{t,l} x_{\text{W}_i,p}^l},&&  \nonumber 
\end{flalign}
where $\{\mathbf{Z}_{t,l}\}_{l=0}^{99}$ are given by
\begin{gather}
\mathbf{Z}_{t,l}=\mathbf{A}_l(\mathbf{B}_{1,1} y_t+\mathbf{B}_{1,2} y_t^{2}+\mathbf{B}_{2,1} y_{4}+\mathbf{B}_{2,2} y_{4}^{2}), \\ \forall l\in [0:99].\nonumber  
\end{gather}

Since the evaluating points $y_{t}$ and $y_{4}$ are identical across the workers in the group $G_t$, we can write the polynomial of $x$ to be decoded by the master as $\mathbf{\tilde A}(x)(\mathbf{\tilde B}_1(y_t)+\mathbf{\tilde B}_2(y_4))=\sum_{l=0}^{99}{\mathbf{Z}_{t,l} x^l}$, whose coefficients are $\{\mathbf{Z}_{t,i}\}_{i=0}^{99}$.
Since the degree of polynomial $\mathbf{\tilde A}(x)(\mathbf{\tilde B}_1(y_t)+\mathbf{\tilde B}_2(y_4))$ is 99 and the evaluating points $\{x_{\text{W}_i,p}\}_{(p,i)=(1,1)}^{(100,12)}$ are distinct from each other, the master can decode the polynomial from 100 sub-computation results returned by the workers in $G_t$, by polynomial interpolation with respect to the variable $x$.

We denote the decoding function by $d_D$ and the $i$th sub-computation result returned to the master within group $G_t$ by $S_{t,i}$.
The master can decode each of the three polynomials $\{\mathbf{\tilde A}(x)(\mathbf{\tilde B}_1(y_t)+\mathbf{\tilde B}_2(y_4))\}_{t=1}^3$ from each set of 100 sub-computation results $\{S_{t,[100]}\}_{t=1}^3$ returned by each of 3 groups $\{G_t\}_{t=1}^3$, so that the coefficients of the polynomials $\{\mathbf{Z}_{t,[0:99]}\}_{t=1}^{3}$ are obtained.
For $l\in[0:99]$, each set of coefficients $\{\mathbf{Z}_{t,l}\}_{t=1}^3$  are evaluations of polynomials $(\mathbf{A}_l\mathbf{B}_{2,1} y_4+\mathbf{A}_l\mathbf{B}_{2,2} y_4^2)+\mathbf{A}_l\mathbf{B}_{1,1} y+\mathbf{A}_l\mathbf{B}_{1,2} y^2$.
Note that the terms $\{\mathbf{A}_l\mathbf{B}_{2,1} y_4+\mathbf{A}_l\mathbf{B}_{2,2} y_4^2\}_{l=0}^{99}$ are constant terms in each polynomial.
For each $l$, the master can decode the polynomial $(\mathbf{A}_l\mathbf{B}_{2,1} y_4+\mathbf{A}_l\mathbf{B}_{2,2} y_4^2)+\mathbf{A}_l\mathbf{B}_{1,1} y+\mathbf{A}_l\mathbf{B}_{1,2} y^2$ from $\{\mathbf{Z}_{t,l}\}_{t=1}^3$ with polynomial interpolation with respect to $y$ so that the coefficients $\{\mathbf{A}_l\mathbf{B}_{1,i}\}_{i=1}^{2}$ are obtained.
After decoding all of the polynomials $\{(\mathbf{A}_l\mathbf{B}_{2,1} y_4+\mathbf{A}_l\mathbf{B}_{2,2} y_4^2)+\mathbf{A}_l\mathbf{B}_{1,1} y+\mathbf{A}_l\mathbf{B}_{1,2} y^2\}_{l=0}^{99}$, the whole computation $\mathbf{AB}_1$ given in (\ref{wholecomputation}) can be recovered from the coefficients $\{\mathbf{A}_l\mathbf{B}_{1,i}\}_{(l,i)=(0,1)}^{(99,2)}$.
Therefore, $K$ equals 300 and  $d_D(S_{1,[100]}, S_{2,[100]}, S_{3,[100]})=\mathbf{AB}_1$.

In the second example, the computational complexity for each sub-computation $\mathbf{\tilde A}(x_{\text{W}_{i},p})(\mathbf{\tilde B}_1(y_t)+\mathbf{\tilde B}_2(y_4))$ is 50 times smaller than that of the sub-computation in the first example, thus implying that even slow workers may return some sub-computations to the master.
That is, whereas the computations of slow workers are ignored in the first example, even slow workers can contribute to overall computation in the second example by partitioning matrix $\mathbf{A}$ into substantially smaller submatrices.
Furthermore, as each worker computes 100 sub-computation results in the second example, the faster workers can compute and return more sub-computation results than the slower workers whereas even the fastest worker may return one sub-computation result at most in the first example.
That is, in the second example, the faster workers are exploited more efficiently than in the first example.  
As a result, the computation time of the second example is reduced compared to the first example.

Although the computation time is reduced in the second example, more communication load for sending the evaluations of $\mathbf{A}$ from the master to the workers is required in the second example. 
In the first example, the master sends one evaluation $\mathbf{\tilde A}(x_{\text{W}_i}) \in \mathbb{F}_q^{r/2\times s}$ to each worker $\text{W}_i$ whereas the master sends 100 evaluations $\{\mathbf{\tilde A}(x_{\text{W}_i,p})\}_{p=1}^{100}$ to each worker $\text{W}_i$, where $\mathbf{\tilde A}(x_{\text{W}_i,p}) \in \mathbb{F}_q^{r/100\times s}$, in the second example.
That is, twice more communication load from the master to the workers is required in the second example.
As a result, we can say that there is a trade-off between communication load and computation time.
Considering both communication load and computation time, the scheme in the first example may outperform the scheme in the second example if the main bottleneck of the coded computation process is communication between the master and the workers.  

We intuitively explain how the master's privacy is protected in both of the examples.
As we assumed, none of the workers collude with each other, which implies that the worker do not know about the evaluating points of other workers.
Therefore, evaluating $\mathbf{\tilde B}_1$ with $y_t$ in group $G_t$ and $\mathbf{\tilde B}_2$ with $y_4$ across workers does not hurt the master's privacy.
For each worker $\text{W}_i$ in $G_t$, we also assumed that the evaluating points $y_t$ and $y_4$ are randomly chosen and distinct from each other, which implies that $\text{W}_i$ cannot identify that the desired matrix is $\mathbf{B}_1$ by the evaluating points $y_t$ and $y_4$.
For encoding the library, since all of the encoded elements in the library are symmetrically summed in to one equation $\mathbf{\tilde B}_1(y_t)+\mathbf{\tilde B}_2(y_4)$, encoding the library does not hurt the master's privacy.
Therefore, each worker $\text{W}_i$ in each group $G_t$ cannot identify that the desired matrix is $\mathbf{B}_1$ throughout the coded computation process, which implies that the master's privacy has been protected across workers.

\subsection{General description}
\label{general}
In this section, we generally describe the private polynomial codes for matrix multiplication.
There are $N$ non-colluding workers $\{\text{W}_{n}\}_{n=1}^{N}$ and each worker has a library $\mathbf{B}$ of $M$ matrices $\{\mathbf{B}_{k}\}_{k=1}^{M}$ where each $\mathbf{B}_k\in  \mathbb{F}_q^{s\times t}$ for sufficiently large finite field $\mathbb{F}_q$. 
The master has a matrix $\mathbf{A}\in \mathbb{F}_q^{r\times s}$ and desires to multiply $\mathbf{A}$ by one of $\{\mathbf{B}_{k}\}_{k=1}^{M}$ in the library $\mathbf{B}$ while keeping the index of desired matrix $\mathbf{B}_{D}$ from all of the workers. 
Matrix $\mathbf{A}$ can be partitioned into $m$ submatrices $\{\mathbf{A}_{k}\}_{k=0}^{m-1}\in \mathbb{F}_q^{r/m\times s}$ and each $\mathbf{B}_k$ can be partitioned into $n-1$ submatrices $\{\mathbf{B}_{k,l}\}_{l=1}^{n-1}\in \mathbb{F}_q^{s\times t/(n-1)}$, where $m,n\in \mathbb{N}^+$. 
The whole computation $\mathbf{AB}_D$ that the master wants to recover is given by
\begin{align}
\mathbf{AB}_D=\begin{bmatrix}
\mathbf{A}_0\mathbf{B}_{D,1} & \mathbf{A}_0\mathbf{B}_{D,2} & \cdots & \mathbf{A}_0\mathbf{B}_{D,n-1}  \\ \mathbf{A}_1\mathbf{B}_{D,1} & \mathbf{A}_1\mathbf{B}_{D,2} & \cdots & \mathbf{A}_1\mathbf{B}_{D,n-1} \\
\vdots & \vdots & \cdots & \vdots \\
\mathbf{A}_{m-1}\mathbf{B}_{D,1} & \mathbf{A}_{m-1}\mathbf{B}_{D,2} & \cdots & \mathbf{A}_{m-1}\mathbf{B}_{D,n-1} 
\end{bmatrix}.
\nonumber
\end{align}

The polynomial codes for $\mathbf{A}$ and $\{\mathbf{B}_{k}\}_{k=1}^M$ are given as follows.
\begin{align}
\mathbf{\tilde A}(x)=\sum_{l=0}^{m-1}\mathbf{A}_l x^{l}, \:\:\:\: \mathbf{\tilde B}_k(y)=\sum_{l=1}^{n-1}\mathbf{B}_{k,l} y^{l},
\label{encoding}
\end{align}
where $k\in[M]$ and $x,y\in \mathbb{F}_q$ for sufficiently large finite field $\mathbb{F}_q$. 

Each sub-computation result to be returned to the master is in the form of $\mathbf{\tilde A}(x_p)(\mathbf{\tilde B}_D(y_{q})+\sum_{k\in [M]\setminus D}{\mathbf{\tilde B}_k(y_{j_k}))}$, for randomly chosen points $x=x_{p}$, $y=y_{q}$, and $\{y_{j_k} | k\in [M]\setminus D\}$, where the points $y=y_{q}$ and $\{y_{j_k} | k\in [M]\setminus D\}$ are distinct from each other. 
For given $y_{q}$, after $m$ sub-computation results $\{\mathbf{\tilde A}(x_p)(\mathbf{\tilde B}_D(y_{q})+\sum_{k\in [M]\setminus D}{\mathbf{\tilde B}_k(y_{j_k}))}\}_{p=1}^m$ are returned to the master, the master can decode the polynomial given by
\begin{flalign}
&\mathbf{\tilde A}(x)(\mathbf{\tilde B}_D(y_{q})+\sum_{k\in [M]\setminus D}{\mathbf{\tilde B}_k(y_{j_k})}) \nonumber&& \\ &=\sum_{l=0}^{m-1}(\mathbf{A}_l\mathbf{\tilde B}_D(y_{q})+\sum_{k\in [M]\setminus D}{\mathbf{A}_l\mathbf{\tilde B}_k(y_{j_k})})x^{l}. \nonumber 
\end{flalign}

The coefficients of the polynomial $\{\mathbf{A}_{l}(\mathbf{\tilde B}_D(y_q)+\sum_{k\in [M]\setminus D}{\mathbf{\tilde B}_k(y_{j_k})})\}_{l=0}^{m-1}$ are obtained by polynomial interpolation for variable $x$. 
After decoding $n$ different polynomials $\{\mathbf{\tilde A}(x)(\mathbf{\tilde B}_D(y_q)+\sum_{k\in [M]\setminus D}{\mathbf{\tilde B}_k(y_{j_k})})\}_{q=1}^n$ for variable $x$ and obtaining the coefficients of the polynomials $\{\mathbf{A}_{l}(\mathbf{\tilde B}_D(y_q)+\sum_{k\in [M]\setminus D}{\mathbf{\tilde B}_k(y_{j_k})})\}_{(l,q)=(0,1)}^{(m-1,n)}$, the master can decode $m$ different polynomials given by
\begin{flalign}
&\{\mathbf{A}_{l}(\mathbf{\tilde B}_D(y)+\sum_{k\in [M]\setminus D}{\mathbf{\tilde B}_k(y_{j_k})})\}_{l=0}^{m-1} \nonumber&& \\ &=\{\sum_{r=1}^{n-1}\mathbf{A}_{l}\mathbf{B}_{D,r} y^{r}+\sum_{k\in [M]\setminus D}{\mathbf{A}_{l}\mathbf{\tilde B}_k(y_{j_k})}\}_{l=0}^{m-1}. \nonumber 
\end{flalign}

By polynomial interpolation for variable $y$, the coefficients of polynomials $\{\mathbf{A}_{l}\mathbf{B}_{D,r}\}_{(l,r)=(0,1)}^{(m-1,n-1)}$ are obtained, from which the whole computation $\mathbf{AB}_D$ can be recovered.
Note that the terms $\{\sum_{k\in [M]\setminus D}{\mathbf{A}_l\mathbf{\tilde B}_k(y_{j_k})\}_{l=0}^{m-1}}$ are constant terms in each of the polynomials $\{\mathbf{A}_{l}(\mathbf{\tilde B}_D(y)+\sum_{k\in [M]\setminus D}{\mathbf{\tilde B}_k(y_{j_k})})\}_{l=0}^{m-1}$.
Therefore, after $mn$ sub-computation results $\{\mathbf{\tilde A}(x_{p})(\mathbf{\tilde B}_D(y_q)+\sum_{k\in [M]\setminus D}{\mathbf{\tilde B}_k(y_{j_k})})\}_{(p,q)=(1,1)}^{(m,n)}$ are returned to the master, the master can recover the whole computation $\mathbf{AB}_D$ through twofold decoding. 
That is, the minimum number  ($K$) of sub-computation results to recover the whole computation, equals $mn$ in private polynomial codes.

We denote the number of evaluations of $\mathbf{\tilde A}$ sent from the master to each worker $\text{W}_i$ by $L$, where $L$ distinct and randomly chosen evaluating points for $\mathbf{\tilde A}$ are given by $\{x_{\text{W}_i,p}\}_{p=1}^{L}$, and $L$ evaluations sent to $\text{W}_i$ are given by $\{\mathbf{\tilde A}(x_{\text{W}_i,p})\}_{p=1}^{L}$.
In Section \ref{example}, $L=1$ in the first example whereas $L=100$ in the second example.
Since the master encodes $\mathbf{A}$ with encoding function $g^D_{\mathbf{A}, \text{W}_i}$,  $g^D_{\mathbf{A}, \text{W}_i}(\mathbf{A})=\{\mathbf{\tilde A}(x_{\text{W}_i,p})\}_{p=1}^{L}$.
The master sends not only evaluations $\{\mathbf{\tilde A}(x_{\text{W}_i,p})\}_{p=1}^{L}$ but also queries $Q_i^D$ for requesting $\text{W}_i$ to encode the library $\mathbf{B}$ with an encoding function $g^D_{\mathbf{B},\text{W}_i}$ and compute the function $f_{\text{W}_i}^D(g^D_{\mathbf{A},\text{W}_i}(\mathbf{A}),g^D_{\mathbf{B},\text{W}_i}(\mathbf{B}))$.
 
Whereas $L$ evaluations of $\mathbf{\tilde A}$ are assigned to $\text{W}_i$, the master can request each worker $\text{W}_i$ to evaluate each $\{\mathbf{\tilde B}_{k}\}_{k=1}^{M}$ at only one point.
This is because $M-1$ distinct and randomly chosen evaluating points $\{y_{j_k}|k\in[M]\setminus D\}$ for $\{\mathbf{\tilde B}_k|k\in[M]\setminus D\}$ should be identical across the workers so that the term $\sum_{k\in [M]\setminus D}{\mathbf{\tilde B}_k(y_{j_k})}$ can be constant.
Otherwise, the master cannot decode the polynomials $\{\mathbf{A}_{l}(\mathbf{\tilde B}_D(y)+\sum_{k\in [M]\setminus D}{\mathbf{\tilde B}_k(y_{j_k})})\}_{l=0}^{m-1}$ by polynomial interpolation for variable $y$.
For each worker $\text{W}_i$, since each $\{\mathbf{\tilde B}_k|k\in[M]\setminus D\}$ is only evaluated at one point $y_{j_k}$, $\mathbf{\tilde B}_D$ should also only be evaluated at one point, say $y_i$, for protecting the master's privacy.
If $\mathbf{\tilde B}_D$ is evaluated at more than one point while each of $\{\mathbf{\tilde B}_k|k\in[M]\setminus D\}$ is only evaluated at one point, the worker $\text{W}_i$ will notice that $\mathbf{B}_D$ is different from the other matrices $\{\mathbf{B}_k|k\in[M]\setminus D\}$ and suspect that $\mathbf{B}_D$ is the desired matrix.

After evaluating points for $\{\mathbf{\tilde B}_{k}\}_{k=1}^{M}$ are determined, each worker $\text{W}_i$ encodes the library $\mathbf{B}$ into $\mathbf{\tilde B}_D(y_i)+\sum_{k\in [M]\setminus D}{\mathbf{\tilde B}_k(y_{j_k})}$.
That is, $g^D_{\mathbf{B},\text{W}_i}(\mathbf{B})=\mathbf{\tilde B}_D(y_i)+\sum_{k\in [M]\setminus D}{\mathbf{\tilde B}_k(y_{j_k})}$.
After encoding $\mathbf{B}$, each worker $\text{W}_i$ computes the function $f_{\text{W}_i}^D(g^D_{\mathbf{A},\text{W}_i}(\mathbf{A}),g^D_{\mathbf{B},\text{W}_i}(\mathbf{B}))=\{\mathbf{\tilde A}(x_{\text{W}_{i},p})(\mathbf{\tilde B}_D(y_i)+\sum_{k\in [M]\setminus D}{\mathbf{\tilde B}_k(y_{j_k})})\}_{p=1}^{L}$.

Although the master can request each worker $\text{W}_i$ to evaluate each $\{\mathbf{\tilde B}_{k}\}_{k=1}^{M}$ at only one point, the master can decode each of the polynomials $\{\mathbf{A}_{l}(\mathbf{\tilde B}_D(y)+\sum_{k\in [M]\setminus D}{\mathbf{\tilde B}_k(y_{j_k})})\}_{l=0}^{m-1}$, whose degree is $n-1$, from at least $n$ evaluations, where $n\geq2$ from (\ref{encoding}). 
Therefore, the master divides the workers into $n$ equal-sized groups according to the evaluating point for $\mathbf{\tilde B}_D$.
Since the groups are equal-sized, there are $N/n$ workers in each group. 
We denote $n$ groups by 
$\{G_t\}_{t=1}^n$ 
and the evaluating point for $\mathbf{\tilde B}_D$ of each $G_t$ by $y=y_{t}$, where 
the points $\{y_{t}\}_{t=1}^n$ are randomly chosen and distinct from each other. 
That is, each worker $\text{W}_i$ in group $G_t$ encodes the library $\mathbf{B}$ into $g^D_{\mathbf{B},\text{W}_i}(\mathbf{B})=\mathbf{\tilde B}_D(y_{t})+\sum_{k\in [M]\setminus D}{\mathbf{\tilde B}_k(y_{j_k})}$ and computes a function $f^D_{\text{W}_i}(g^D_{\mathbf{A}, \text{W}_i}(\mathbf{A}), g^D_{\mathbf{B},\text{W}_i}(\mathbf{B}))=\{\mathbf{\tilde A}(x_{\text{W}_i,p})(\mathbf{\tilde B}_D(y_{t})+\sum_{k\in [M]\setminus D}{\mathbf{\tilde B}_k(y_{j_k})})\}_{p=1}^L$. 
After $m$ sub-computation results 
$\{\mathbf{\tilde A}(x_{p})(\mathbf{\tilde B}_D(y_t)+\sum_{k\in [M]\setminus D}{\mathbf{\tilde B}_k(y_{j_k})})\}_{p=1}^m$ 
are returned to the master across the workers in each group $G_t$, the master can decode the polynomial 
$\mathbf{\tilde A}(x)(\mathbf{\tilde B}_D(y_t)+\sum_{k\in [M]\setminus D}{\mathbf{\tilde B}_k(y_{j_k})})$ 
for variable $x$.
After each of $n$ groups returns $m$ sub-computation results, the master can decode $m$ different polynomials  
$\{\mathbf{A}_{l}(\mathbf{\tilde B}_D(y)+\sum_{k\in [M]\setminus D}{\mathbf{\tilde B}_k(y_{j_k})})\}_{l=0}^{m-1}$ 
for variable $y$ 
so that the decoding results are given by $\{\mathbf{A}_{l}\mathbf{B}_{D,r}\}_{(l,r)=(1,1)}^{(m,n-1)}$,
from which the whole computation $\mathbf{AB}_D$ is recovered. 

Let us examine the relation between the parameters.
For $m$ and $L$, $L\leq m$ because $L=m$ implies that the master sends $m$ evaluations of $\mathbf{\tilde A}$ to each worker, where the size of each evaluation is $1/m$ of whole matrix $\mathbf{A}$.
That is, if $L=m$, the whole information of $\mathbf{A}$ is sent to each worker.
If we denote the communication load for sending whole matrix $\mathbf{A}$ from the master to each worker by $|\mathbf{A}|$, the communication load from the master to each worker in private polynomial code is given by $\frac{L}{m}|\mathbf{A}|$.

For $m$. $n$, $N$, and $L$, in order for the master to decode the polynomial $\mathbf{\tilde A}(x)(\mathbf{\tilde B}_D(y_t)+\sum_{k\in [M]\setminus D}{\mathbf{\tilde B}_k(y_{j_k})})$ from $m$ sub-computation results returned by $G_t$, the maximum number of sub-computation results that can be returned by $G_t$, which is given by $L\frac{N}{n}$, should not be smaller than $m$.
We can express this relation as follows.
\begin{align}
L\frac{N}{n}\geq m.
\label{LNmn}
\end{align}

\subsection{Privacy proof}
\label{proof}

In this section, we show that the master's privacy is protected in private polynomial codes.
In particular, we show that the privacy constraint for each worker $\text{W}_i$ in group $G_t$ is satisfied, which was given by (\ref{privacyconstraint}).
By chain rule, we can write the privacy constraint as follows.
\begin{flalign}
&I(D;Q_i^D,C_i^D,R_i^D,\mathbf{B}) \nonumber\\
&=I(D;Q_i^D) \nonumber\\
&+I(D;\mathbf{B}|Q_i^D) \nonumber\\
&+I(D;C_i^D|Q_i^D,\mathbf{B}) \nonumber \\
&+I(D;R_i^D|Q_i^D,\mathbf{B},C_i^D) \nonumber
\end{flalign}

Note that $R_i^D=f^D_{\text{W}_i}(g^D_{\mathbf{A},\text{W}_i}(\mathbf{A}), g^D_{\mathbf{B},\text{W}_i}(\mathbf{B}))$ is a deterministic function of $C_i^D=g^D_{\mathbf{A},\text{W}_i}(\mathbf{A})$ and $g^D_{\mathbf{B},\text{W}_i}(\mathbf{B})$, where $g^D_{\mathbf{B},\text{W}_i}(\mathbf{B})$ is a function of $\mathbf{B}$. 
Since $g^D_{\mathbf{B},\text{W}_i}(\mathbf{B})=\mathbf{\tilde B}_D(y_i)+\sum_{k\in [M]\setminus D}{\mathbf{\tilde B}_k(y_{j_k})}$ and the evaluating points $y_t$ and $\{y_{j_k}|k\in[M]\setminus D\}$ are determined by the the queries $Q_i^D$, $g^D_{\mathbf{B},\text{W}_i}(\mathbf{B})$ is a deterministic function of $Q_i^D$, which implies that $R_i^D$ is a deterministic function of $C_i^D$, $\mathbf{B}$, and $Q_i^D$.
Therefore, $I(D;R_i^D|Q_i^D,\mathbf{B},C_i^D)=0$. 
Since $C_i^D=\{\mathbf{\tilde A}(x_{\text{W}_i,p})\}_{p=1}^L=\{\sum_{l=0}^{m-1}{\mathbf{A}_lx_{\text{W},p}^l}\}_{p=1}^L$, $C_i^D$ is independent of $D$. 
Therefore, $I(D;C_i^D|Q_i^D,\mathbf{B})=0$. 
Since the master determines the index of the desired matrix $D$ without knowing any information of the library $\mathbf{B}$, the library $\mathbf{B}$ is independent of $D$, which is followed by $I(D;\mathbf{B}|Q_i^D)=0$. 

For the desired matrix $\mathbf{B}_{D}$, the master sends queries $Q_i^D$ to each worker $\text{W}_i$ in $G_t$ in order to request $\text{W}_i$ to encode the library $\mathbf{B}$ into $\mathbf{\tilde B}_D(y_t)+\sum_{k\in [M]\setminus D}{\mathbf{\tilde B}_k(y_{j_k})}$ and compute the function $\{\mathbf{\tilde A}(x_{\text{W}_i,p})(\mathbf{\tilde B}_D(y_t)+\sum_{k\in [M]\setminus D}{\mathbf{\tilde B}_k(y_{j_k})})\}_{p=1}^{L}$.
The queries $Q_i^D$ are fourfold:
\begin{enumerate}
    \item $Q_{i,p}^D$ : queries for partitioning each matrix $\mathbf{B}_{k}$ in the library $\mathbf{B}$ into $n-1$ submatrices $\{\mathbf{B}_{k,l}\}_{l=1}^{n-1}$
    \item $Q_{i,e}^D$ : queries for evaluating $\mathbf{\tilde B}_{D}$ and $\{\mathbf{\tilde B}_{k}|k\in [M]\setminus D\}$ at the points $y_t$ and $\{y_{j_k}|k\in[M]\setminus D\}$, respectively
    \item $Q_{i,s}^D$ : queries for summing the evaluations of $\{\mathbf{\tilde B}_{k}\}_{k=1}^M$ into one equation $\mathbf{\tilde B}_D(y_t)+\sum_{k\in [M]\setminus D}{\mathbf{\tilde B}_k(y_{j_k})}$
    \item $Q_{i,c}^D$ : queries for computing the function
\end{enumerate}

According to $Q_{i,p}^D$, all submatrices $\{\mathbf{B}_{k,l}\}_{(k,l)=(1,1)}^{(M,n-1)}$ are elements in $\mathbb{F}_q^{s\times t/(n-1)}$.
Therefore, $Q_{i,p}^D$ are independent of $D$, which implies $I(D;Q_{i,p}^D)=0$.

According to $Q_{i,e}^D$, as assumed in Section \ref{general}, the points $y_t$ and $\{y_{j_k} | k\in [M]\setminus D\}$ are distinct from each other and randomly chosen in $\mathbb{F}_q$.
Therefore, $Q_{i,e}^D$ are independent of $D$, which implies $I(D;Q_{i,e}^D)=0$.

According to $Q_{i,s}^D$, all of the evaluations of $\{\mathbf{\tilde B}_k\}_{k=1}^M$ are symmetrically summed into one equation $\mathbf{\tilde B}_D(y_t)+\sum_{k\in [M]\setminus D}{\mathbf{\tilde B}_k(y_{j_k})}$.
Therefore, $Q_{i,s}^D$ are independent of $D$, which implies $I(D;Q_{i,s}^D)=0$.

According to $Q_{i,c}^D$, each of $C_i^D=\{\mathbf{\tilde A}(x_{\text{W}_i,p})\}_{p=1}^L$ is multiplied by $\mathbf{\tilde B}_D(y_t)+\sum_{k\in [M]\setminus D}{\mathbf{\tilde B}_k(y_{j_k})}$.
We already explained that $C_i^D$ and $\mathbf{\tilde B}_D(y_t)+\sum_{k\in [M]\setminus D}{\mathbf{\tilde B}_k(y_{j_k})}$ are independent of $D$.
Therefore, $Q_{i,c}^D$ are also independent of $D$, which implies $I(D;Q_{i,c}^D)=0$.

As a result, $I(D;Q_{i}^D)=0$, which implies that $I(D;Q_i^D,C_i^D,R_i^D,\mathbf{B})=0$.
Since the privacy constraint is satisfied for every worker, the master's privacy is considered to be protected in private  polynomial codes. $\square$

\subsection{Special cases}
\label{sp}

In this section, we propose certain special cases of private polynomial codes, namely, private one-shot polynomial codes and private asynchronous polynomial codes.

In private one-shot polynomial codes, the master sends only one evaluation of $\mathbf{\tilde A}$ to each worker so that each worker returns only one sub-computation result to the master, which implies that $L=1$.
The first example in Section \ref{example} corresponds to the private one-shot polynomial code.

In private asynchronous polynomial codes, the master divides its own matrix $\mathbf{A}$ into much smaller partitions, which implies that $m$ becomes much larger in turn.
The second example in Section \ref{example} corresponds to the private asynchronous polynomial codes, where $m=100$.
If $m$ is sufficiently large, even the slow workers can compute and return the sub-computations with their low computational capabilities, which results in improved efficiency of coded computation.
As $m$ becomes larger, the lower bound of $L$ also becomes higher, since $L\geq \frac{mn}{N}$ from (\ref{LNmn}).
As $L$ becomes larger, faster workers can compute and return more sub-computations to the master.
That is, faster workers are exploited more efficiently in the private asynchronous polynomial codes as compared to the private one-shot polynomial codes where the faster workers are not exploited after returning only one sub-computation.
Therefore, by properly designing the parameters $m$ and $L$, all of the workers in each group can continue working until $m$ sub-computation results are returned. 
This implies that asynchronous coded computation is group-wise in private asynchronous polynomial codes, where the definition of asynchronous coded computation was given in Section \ref{intro}.

\section{Simulation results}

In this section, in terms of the computation time consumed for receiving $K$ sub-computation results across $N$ workers and communication load from the master to $N$ workers, we compare the private one-shot polynomial codes and private asynchronous polynomial codes with the conventional private information retrieval (PIR) scheme. 
We benchmark the optimal RPIR scheme in \cite{RPIR_Sun}, where the number of unresponsive nodes corresponds to the stragglers. 
For fair comparison, we assume that the workers do not collude with each other in RPIR scheme, which is the same as in private polynomial codes. 

The RPIR scheme in \cite{RPIR_Sun} is directly applicable to the coded computation.
The master encodes the library $\mathbf{B}=\{\mathbf{B}_k\}_{k=1}^M$ into $\{\mathbf{\tilde B}_k\}_{k=1}^{M}$ with MDS code and each $\mathbf{\tilde B}_k$ is multiplied by  $\mathbf{A}$. 
That is, the master should transmit all of matrix $\mathbf{A}$ to each worker.
Encoding $\mathbf{A}$ does not affect the computation time and the communication load since every encoded matrix $\mathbf{\tilde A}$ should be multiplied across the workers to exploit the encoded undesired matrices $\{\mathbf{\tilde B}_k|k\in [M]\setminus D\}$ as side information. 

\subsection{Computation time}

For computation time, we assume that the computation time distribution of each worker is independent of each other and follows the exponential distribution as in \cite{KLee}. 
That is, if we denote the computation time of a single worker by random variable $T_0$, the following holds.
\begin{align}
\text{Pr}(T_0\leq t)=1-e^{-\mu(t-\gamma)}, \forall t\geq \gamma, \nonumber
\end{align}
where the parameters $\gamma$ and $\mu$ denotes the shift parameter and straggling parameter, respectively. 

When computing $\mathbf{AB}_D$ without considering the master's privacy as in \cite{KLee}, the computation time is that of the $K$th fastest worker for computing $1/K$ of whole computation.
Therefore, the computation time is the expected value of the $K$th statistic of $N$ independent exponential random variables.
If we denote a sum $\sum_{n=1}^N{\frac{1}{n}}$ by $H_N$, $H_N\simeq \log{N}$ for large $N$.
Since the expected value of the $K$th statistic of $N$ independent exponential random variable is given by $\frac{H_N-H_{N-K}}{\mu}$, the computation time of the conventional coded computation is given by 
\begin{align}
t_{conv} = \frac{1}{K}(\gamma + \frac{1}{\mu}\text{log}\frac{N}{N-K}) \nonumber.
\end{align} 

Similar to the conventional coded computation in \cite{KLee}, in the RPIR scheme, the computation of stragglers is ignored and the computation time is determined by the $K$th fastest worker. 
In order to recover $\mathbf{AB}_D$ under the privacy constraint, the RPIR scheme requires $1+\frac{1}{K}+\cdots+\frac{1}{K^{M-1}}$ times more computation than that required by directly computing $\mathbf{AB}_D$ without considering the master's privacy. 
Therefore, the computation time of the RPIR scheme denoted by $t_{RPIR}$ takes $(1+\frac{1}{K}+\cdots+\frac{1}{K^{M-1}})$ times longer than $t_{conv}$, which is given by 
\begin{align}
t_{RPIR}=(\frac{1}{K}+\cdots+\frac{1}{K^{M}})(\gamma + \frac{1}{\mu}\text{log}\frac{N}{N-K}).
\label{tRPIR}
\end{align}

As explained in Section \ref{general}, $K=mn$ in private polynomial codes, while $\mathbf{A}$ and each of $\{\mathbf{B}_k\}_{k=1}^M$ are divided into $m$ and $n-1$ submatrices, respectively.
That is, the sizes of $\mathbf{\tilde A}$ and encoded $\mathbf{B}$ are $\frac{1}{m}$ of $\mathbf{A}$ and $\frac{1}{n-1}$ of each $\{\mathbf{B}_k\}_{k=1}^M$, respectively.
Therefore, the computational complexity for each sub-computation is $\frac{1}{m(n-1)}$ of the computational complexity for directly computing $\mathbf{AB}_D$ without considering the master's privacy.

In order to derive the computation time in the private one-shot polynomial codes and the private asynchronous polynomial codes, we order the workers from the fastest worker to the slowest worker.
That is, worker $\text{W}_i$ is the $i$th fastest worker among the $N$ workers.
As explained in Section \ref{general}, the computation time of private polynomial codes equals the computation time of the slowest group.
We denote the slowest group by $G_s$ where $s\in[n]$.
If we denote the workers in $G_s$ by 
$\{\text{W}_{s_i}\}_{i=1}^{N/n}$, worker $\text{W}_{s_i}$ is the $s_i$th fastest worker among all workers $\{\text{W}_i\}_{i=1}^N$ and we assume $s_1<s_2<\cdots<s_{N/n}$.

Therefore, the computation time of private one-shot polynomial codes equals the computation time for the worker $\text{W}_{s_m}$ to compute its sub-computation, which is given by 

\begin{align}
t_{one}=\frac{1}{m(n-1)}(\gamma + \frac{1}{\mu}\text{log}\frac{N}{N-s_m}).
\label{tone}
\end{align}

Since there are $\frac{N!}{{((N/n)!)}^n}\frac{1}{n!}$
combinations for grouping $N$ workers into $n$ groups, we average $t_{one}$ over all of the combinations so as to obtain average computation time $t_{a,one}$. We denote the set of combinations by $\{T_i\}_{i=1}^{\frac{N!}{{((N/n)!)}^n}\frac{1}{n!}}$ and $t_{one}$ for $T_i$ by $t_{one,i}$, so that $t_{a,one}$ is given by

\begin{align}
t_{a,one}=\frac{\sum_{i=1}^{\frac{N!}{{((N/n)!)}^n}\frac{1}{n!}}{t_{one,i}}} {\frac{N!}{{((N/n)!)}^n}\frac{1}{n!}}. 
\end{align}

For the private asynchronous polynomial codes, since every worker in the slowest group $G_s$ continues working during the computation time $t_{async}$, the following holds.
\begin{flalign}
t_{async} &=P_{s_1}(\gamma + \frac{1}{\mu}\text{log}\frac{N}{N-s_1}) \nonumber \\ \nonumber &=P_{s_2}(\gamma + \frac{1}{\mu}\text{log}\frac{N}{N-s_2}) \\ &\:\:\:\:\:\:\:\:\:\:\:\:\:\:\:\:\:\:\:\:\:\:\:\: \vdots \nonumber \\ &= P_{s_{N/n}}(\gamma + \frac{1}{\mu}\frac{N}{N-s_{N/n}}) \nonumber,
\end{flalign}
where $P_{s_i}$ denotes the normalized amount of sub-computations returned by $\text{W}_{s_i}$ in $G_s$ during $t_{async}$.

Since $\sum_{i=1}^{N/n}P_{s_i}=t_{async}(\sum_{i=1}^{N/n}{(\gamma + \frac{1}{\mu}\text{log}\frac{N}{N-{s_i}}})^{-1})=\frac{1}{n-1}$, $t_{async}$ is given by 
\begin{align}
t_{async}=\frac{1/(n-1)}{\sum_{i=1}^{N/n}{\frac{1}{\gamma + \frac{1}{\mu}\text{log}\frac{N}{N-{s_i}}}}}. 
\end{align}

As with private one-shot polynomial code, we average $t_{async}$ over all combinations of grouping so as to obtain average computation time $t_{a,async}$.
If we denote $t_{async}$ for given combination $T_i$ by $t_{async,i}$,  $t_{a,async}$ is given by
\begin{align}
t_{a,async}=\frac{\sum_{i=1}^{\frac{N!}{{((N/n)!)}^n}\frac{1}{n!}}{t_{async,i}}} {\frac{N!}{{((N/n)!)}^n}\frac{1}{n!}}. 
\label{tasync}
\end{align}

\begin{figure}[t]
    \centerline{\includegraphics[width=9cm]{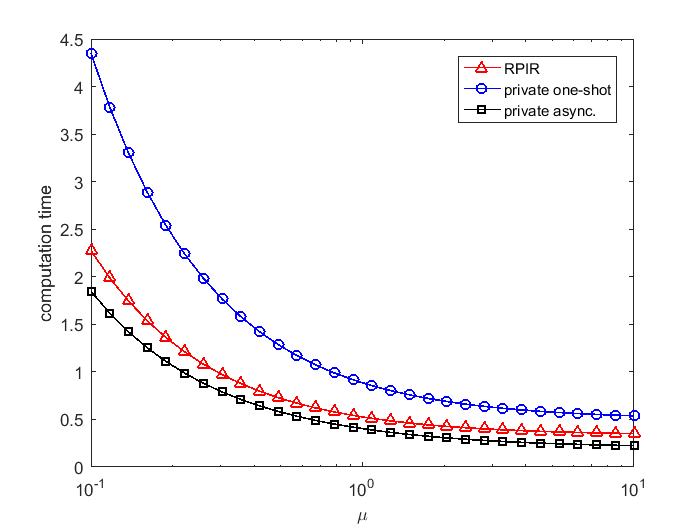}}
    \caption{The computation time comparison between the RPIR scheme, private one-shot polynomial code and private asynchronous polynomial code for $N=12$, $M=4$, $\mu=\gamma=0.1$, and varying $K$.}
    \label{computation}
    \vspace{0mm}
\end{figure}

We compare the computation time between three schemes for $N=12$, $M=4$, and $\gamma=\mu=0.1$. 
For the private one-shot polynomial codes and the private asynchronous polynomial codes, we set $n=2$. 
Since $K=mn$ in the private polynomial codes, we set $K$ as even number and vary $K$ from $2$ to $10$.
That is, $m$ is varying from 1 to 5 in the private one-shot polynomial codes.
Note that $t_{a,async}$ is independent of $m$ from (\ref{tasync}).
The comparison result for computation time is given in Fig. \ref{computation}. 
The computation time of private one-shot polynomial cod is strictly larger than that of the RPIR scheme. 
The computation time of private asynchronous polynomial codes is given by 1.5861, where the private asynchronous polynomial codes achieve at least 60\% and 20\% reduction in computation time for given parameters, compared to the private one-shot polynomial codes and RPIR scheme, respectively.

\begin{figure}[t]
    \centerline{\includegraphics[width=9cm]{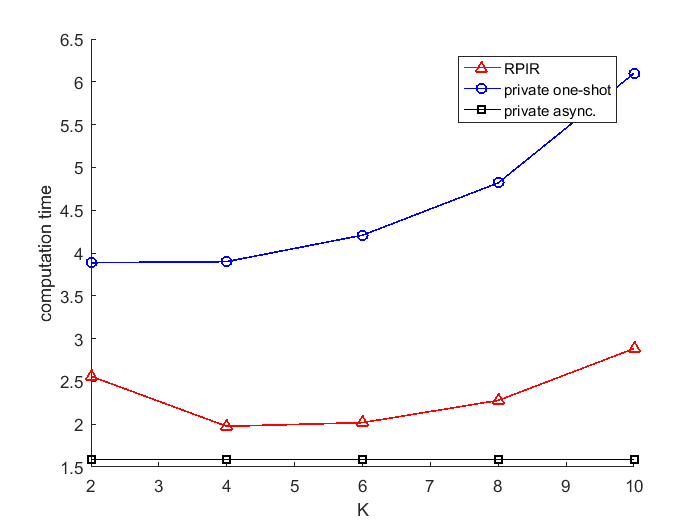}}
    \caption{The computation time comparison between the RPIR scheme, private one-shot polynomial code and private asynchronous polynomial code for $N=12$, $M=4$, $K=4$, $\gamma=1$, and varying $\mu$ from $10^{-1}$ to $10$.}
    \label{computation2}
    \vspace{0mm}
\end{figure}

The shift parameter $\gamma$ and straggling parameter $\mu$ also affect the computation time. 
In order to identify the effects of the two parameters, we compare the computation times between the three schemes for varying $\mu$ from $10^{-1}$ to $10$ and  $\gamma=1$. 
For $N=12$, $M=4$, we set $K=4$ and $n=2$. 
The comparison result is given in Fig. \ref{computation2}. 
The private asynchronous polynomial code outperforms the other schemes, although the gap is diminished as $\mu$ increases. 
This is reasonable because small $\mu$ indicates that the delaying effect by the slow workers is significant, where the delaying effect is alleviated in private asynchronous polynomial codes, since the computation of slow workers is not ignored. 
As $\mu$ becomes larger, the delaying effect of slow workers becomes negligible, thus implying that the advantage of private asynchronous polynomial code also becomes negligible. 
Although the performance gap is reduced as $\mu$ increases, the private one-shot polynomial code is always the worst in the entire range.

\subsection{Communication load}

Let us characterize the communication load of each scheme. 
We only consider the communication load from the master to the workers.
In private polynomial codes, the communication load from the workers to the master is identical regardless of the parameter $m$ or $L$.
Note that each sub-computation in the private polynomial codes is an element in $F_q^{\frac{r}{m}\times \frac{t}{(n-1)}}$, whereas $mn$ sub-computations are returned to the master, thus implying that the communication load for returning $mn$ sub-computations only depends on $n$.
Therefore, for given $n$, the communication loads of private one-shot polynomial codes and private asynchronous polynomial codes are identical.
Therefore, we compare only communication load from the master to the workers.

We assume that the master simultaneously transmits its own data to every worker.
Since all of $\mathbf{A}$ should be transmitted to every worker in the RPIR scheme, the communication load is $N|\mathbf{A}|$. 
In private one-shot polynomial codes, the size of $\mathbf{\tilde A}$ is $|\mathbf{A}|/m$ and one evaluation of $\mathbf{\tilde A}$ is delivered to each worker, which implies that $L=1$. 
Therefore, the communication load of the private one-shot polynomial codes is $\frac{N}{m}|\mathbf{A}|$. 
From (\ref{LNmn}), $m$ is upper bounded by $N/n$ in the private one-shot polynomial code.
Therefore, whereas $t_{a,one}$ was strictly larger than $t_{RPIR}$, the communication load of the RPIR scheme is at most $N/n$ times larger than that of the private one-shot polynomial codes.

Since the master sends $L$ evaluations of $\mathbf{\tilde A}$ to each worker in private asynchronous polynomial codes, the communication load in private asynchronous polynomial code is given by $\frac{NL}{m}|\mathbf{A}|$. 
In order to cover the extreme case in which the fastest worker in a group solely returns all of $m$ sub-computations before the other workers return any sub-computation, we set $L=m$, which is the same as in the second example in Section \ref{example}.
In this case, the communication load of private asynchronous polynomial codes becomes $N|\mathbf{A}|$, which is the same as that of the RPIR scheme. 
However, for many practical scenarios, $L$ can be significantly lowered in private asynchronous polynomial codes, thus implying that the gap of communication load between the private one-shot polynomial codes and the private asynchronous polynomial codes can be reduced.

Note that if we minimize the communication load of private one-shot polynomial codes by maximizing $m$, the computation time of private one-shot polynomial codes in (\ref{tone}) is maximized since $K$ is fixed to $mn$.
That is, there is a trade-off between computation time and communication load in private one-shot polynomial codes. 

\section{Conclusion}
In this paper, we introduced private coded computation as a variation of coded computation that protects the master's privacy.
As an achievable scheme for private coded computation, we proposed private polynomial codes based on conventional polynomial codes in coded computation. 
As special cases of private polynomial codes, we characterized private one-shot polynomial codes and private asynchronous polynomial codes.
While the private asynchronous polynomial codes achieved faster computation time, the private one-shot polynomial codes achieved smaller communication load.
We compared the two special private polynomial codes with the conventional RPIR scheme and verified that the proposed schemes outperform the conventional scheme either in terms of computation time or communication load. 
In future work, we may use different codes in order to improve the performance. 
For example, \textit{entangled polynomial codes} \cite{entangled}, a generalized version of polynomial codes, can be applied to private polynomial codes.

\begin{IEEEbiography}[{\includegraphics[width=1in,height=1.25in,clip,keepaspectratio]{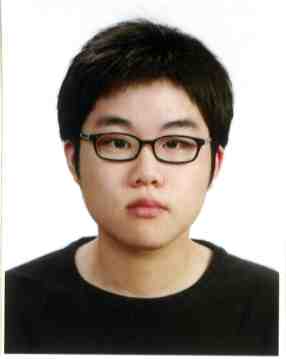}}]{Minchul Kim} (S'17) received the B.S. degree in Electrical and Computer Engineering from Seoul National University, South Korea in 2014. He is currently working towards the Ph.D. degree with the Department of Electrical and Computer Engineering, Seoul National University, South Korea. His research interests include distributed computing, private information retrieval, and codes for distributed storage systems.
\end{IEEEbiography}

\begin{IEEEbiography}[{\includegraphics[width=1in,height=1.25in,clip,keepaspectratio]{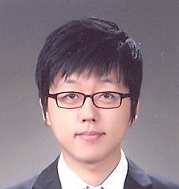}}]{Heecheol Yang} (S'15) received the B.S. degree in Electrical and Computer Engineering from Seoul National University, South Korea in 2013. He is currently working towards the Ph.D. degree with the Department of Electrical and Computer Engineering, Seoul National University, South Korea. His research interests include interference management techniques for wireless communications, network information theory, and security for data storage systems. He received the Bronze prize in the 23rd Samsung Humantech Paper Contest. 
\end{IEEEbiography}

\begin{IEEEbiography}[{\includegraphics[width=1in,height=1.25in,clip,keepaspectratio]{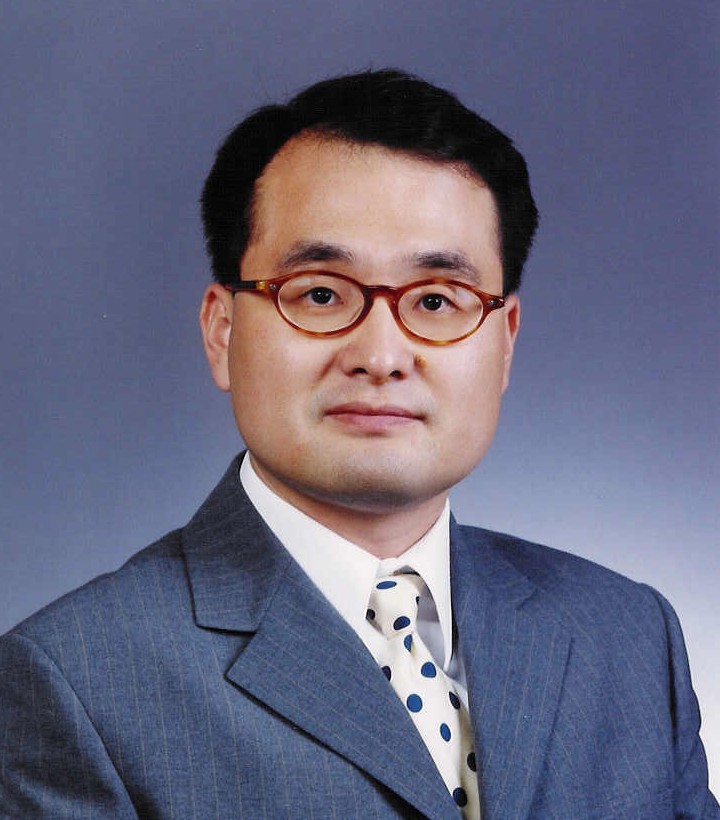}}]{Jungwoo Lee} (S'88, M'94, SM'07) received the B.S. degree in Electronics Engineering from Seoul National University, Seoul, Korea in 1988 and the M.S.E. degree and the Ph.D. degree in Electrical Engineering from Princeton University in 1990 and 1994. 

He is now a professor at the department of electrical and computer engineering of Seoul National University. He was a member of technical staff working on multimedia signal processing at SRI (Sarnoff) from 1994 to 1999, where he was a team leader (PI) for an \$18M NIST ATP program. He has been with Wireless Advanced Technology Lab of Lucent Technologies Bell Labs since 1999, and worked on W-CDMA base station algorithm development as a team leader, for which he received two Bell Labs technical achievement awards. His research interests include wireless communications, information theory, distributed storage, and machine learning. He holds 21 U.S. patents. He is an editor for IEEE wireless communications letters (WCL) since 2017. He was an associate editor for IEEE Transactions on Vehicular Technology (2008 to 2011) and Journal of Communications and Networks (JCN, 2012-2016). He has also been a chief editor for KICS journal, and an executive editor for Elsevier-KICS ICT Express since 2015. He has also been a Track chair for IEEE ICC SPC (2016-2017), and was a TPC/OC member for ICC’15, ITW’15, VTC’15s, ISIT’09, PIMRC’08, ICC’05, and ISITA’05. He received the Qualcomm Dr. Irwin Jacobs award in 2014 for his contributions in wireless communications.
\end{IEEEbiography}

\end{document}